\documentclass[aip, reprint]{revtex4-1}

\usepackage{graphicx}

\def\Rb{$^{87}$Rb }

\begin{document}

\title{Evolutionary optimization of an experimental apparatus}

\author{I. Geisel}
\email[]{Electronic mail: geisel@iqo.uni-hannover.de}
\affiliation{ Institut f\"ur Quantenoptik, Leibniz Universit\"at Hannover, 30167~Hannover, Germany}

\author{ K. Cordes}
\affiliation{ Institut f\"ur Informationsverarbeitung, Leibniz Universit\"at Hannover, 30167~Hannover, Germany}
\author{J. Mahnke}\affiliation{ Institut f\"ur Quantenoptik, Leibniz Universit\"at Hannover, 30167~Hannover, Germany}

\author{S. J\"ollenbeck}\affiliation{ Institut f\"ur Quantenoptik, Leibniz Universit\"at Hannover, 30167~Hannover, Germany}
\author{ J. Ostermann}
\affiliation{ Institut f\"ur Informationsverarbeitung, Leibniz Universit\"at Hannover, 30167~Hannover, Germany}
\author{ J. Arlt}
\affiliation{ QUANTOP, Danish National Research Foundation Center for Quantum Optics, Aarhus University, 8000~Aarhus, Denmark}
\author{W. Ertmer}
\author{C. Klempt}\affiliation{ Institut f\"ur Quantenoptik, Leibniz Universit\"at Hannover, 30167~Hannover, Germany}

\date{\today}

\begin{abstract}

In recent decades, cold atom experiments have become increasingly complex. While computers control most parameters, optimization is mostly done manually. This is a time-consuming task for a high-dimensional parameter space with unknown correlations. Here we automate this process using a genetic algorithm based on Differential Evolution. We demonstrate that this algorithm optimizes 21 correlated parameters and that it is robust against local maxima and experimental noise. The algorithm is flexible and easy to implement. Thus, the presented scheme can be applied to a wide range of experimental optimization tasks.
\end{abstract}

\maketitle

Due to the progress in digital electronics and information technology, a wide range of experimental parameters in modern physics experiments is controlled by computer software. Usually, it is necessary to adjust a large number of parameters for the proper operation of the experimental apparatus. The parameters are determined according to a specific optimization criterion. This optimization task is ubiquitous to experiments in all fields of physics, including the setup and adjustment of accelerator\cite{Hajima1992, Bacci2007} and decelerator\cite{Gilijamse2006} beam lines, automated scanning probe microscopy in  solid state physics\cite{Woolley2011}, wing design for supersonic transportation\cite{Obayashi2000}, and the shaping of ultrashort laser pulses\cite{Zeidler2001, Assion1998}. The field of ultracold quantum gases is a prominent example, where computer control systems with many adjustable parameters are required, since the recent development of a wide range of preparation and manipulation tools has led to increasingly complex setups and experimental protocols. 

 \begin{figure}[h!tb]
		\includegraphics[width={8.5 cm}]{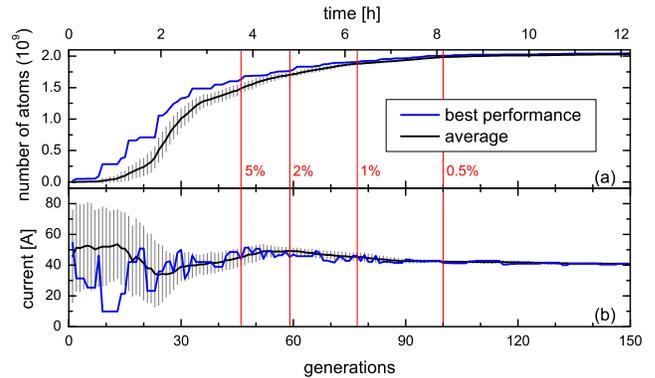}
		\caption{Example for a 21-dimensional optimization. We have optimized 21 experimental parameters in a cold atom experiment. The vertical lines indicate, when certain thresholds of the termination criterion are met. The optimization met the $0.5\%$-termination condition after 100 generations or 8 hours 10 minutes. (a) The objective function was defined by the final number of trapped atoms. The plot shows the best (blue) and the average (black) value of each generation. The error bars indicate the standard deviation within the generation. (b) As an example, the optimization of one of the parameters (a current) is demonstrated by plotting the value of the best performing individual (blue) and the generation average (black). In total, the example illustrates how the algorithm samples a decreasing subspace of the parameter space until the objective function converges to the global maximum and optimal parameter values are found.}
	\label{fig:21D}
\end{figure}

In many cases, the experimental parameters are optimized manually. This is only feasible if most parameters can be optimized independently. The manual optimization of correlated parameters involves an incremental search on a multi-dimensional parameter space. The effort for such a grid-based optimization grows exponentially with the number of parameters. Therefore, the grid-based optimization of a large number of correlated parameters becomes practically impossible. High-dimensional optimization tasks require an automatic solution which scales better than exponential. Up to now, only one global optimization of a cold atom experiment has been demonstrated~\cite{Rohringer2008, Rohringer2011}. However, the demonstration was limited to four correlated parameters. A global optimization of a large number of correlated parameters has not been presented. 

In this Letter, we present the global optimization of up to 21 correlated parameters in a cold atom experiment (see Fig.~\ref{fig:21D}). The proposed heuristic optimization is based on an algorithm called \textit{Differential Evolution} (DE), which was invented in 1995~\cite{Storn1995}. It has been tested in a wide range of applications, primarily in computer science~\cite{Price2005,Cordes2009}. 
In Ref.~\cite{Jin2005,Das2005} DE is extended to handle noisy environments. While this work is mainly restricted to resampling and averaging, we propose an algorithm which is especially robust to technical noise and adapts to time-varying experimental conditions. We call this algorithm \textit{Limited Individual Lifetime Differential Evolution} (LILDE). We demonstrate that the optimization time of LILDE scales better than quadratically with the number of parameters. These advantages make LILDE an ideal optimization strategy for experiments with a large number of correlated parameters. It has possible applications in all fields of experimental physics, and, in general, for optimization tasks of a noisy objective function.

\begin{figure}[h!tb]
\includegraphics[width={6.34cm}]{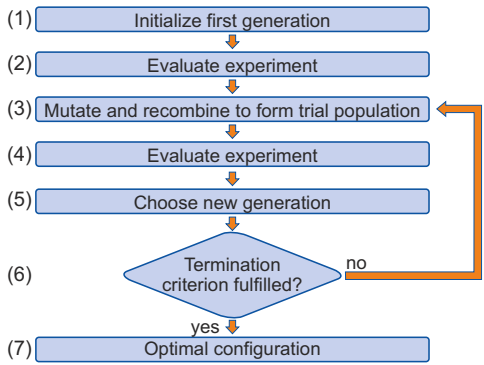}
\caption{The process of the algorithm. (1) The initial generation is randomly distributed over the parameter space. (2) The experiment is run for each candidate of the initial generation to  evaluate the objective function. (3) The current generation is mutated to form a trial population. (4) The experiment is run for each candidate of the trial population. (5) If a better results is achieved in the experiment, the new candidate replaces the former. (6) The process is repeated until the current generation meets a termination condition. (7) The optimal configuration is reached.}
	\label{fig:flowchart}
\end{figure}

In the following, we summarize the DE algorithm and the proposed extensions. DE finds a global optimum of an objective function in a multidimensional parameter space. In our case, each vector in the parameter space corresponds to a possible experimental setting. The experiment is performed with such a parameter set and the outcome is evaluated with respect to the desired optimization criterion. Such an experimental run represents a single evaluation of the objective function in the parameter space. The operation is based on sets (populations) of such vectors which evolve in generations. Figure~\ref{fig:flowchart} illustrates the operation sequence of the algorithm.

Each generation consists of $N$ vectors with dimension $d$. (1) The initial generation $\{\vec{x_i}| i \in[0,N-1]\}$ is randomly distributed over the parameter space with lower boundaries $L^n$ and upper boundaries $U^n$ and $n \in[0,d-1]$. (2) This first generation is evaluated in the experiment. (3) The trial population $\{\vec{u_i}| i \in[0,N-1] \}$ is formed in two steps. First, a mutant population $\{\vec{v_i}| i \in[0,N-1]\}$ is generated. Second, each mutated vector $\vec{v_i}$ is combined with one vector $\vec{x_i}$ from the previous generation. The resulting vector $\vec{u_i}$ is a new candidate in the trial population.

\textit{Mutation:} Three randomly chosen, mutually exclusive vectors $\vec{x_j}$, $\vec{x_k}$, and $\vec{x_l}$ with $j, k, l \in[0,N-1]$ are combined: 

\begin{equation}
\vec{v}_i= \vec{x}_j+F \cdot (\vec{x}_k-\vec{x}_l), \nonumber
\end{equation}

where $F\in ]0,1]$ is one of the DE constants called amplification constant. The vectors $\vec{x_j}$, $\vec{x_k}$, and $\vec{x_l}$ are not chosen from the entire set but from a subset with the best performance. A second constant $E \in ]0,1]$ called elite parameter defines the relative size of the subset. Each vector component $v_i^n$, that falls outside the parameter space boundaries $L^n$ or $U^n$, is placed on the appropriate boundary.

\textit{Recombination:} Each vector $\vec{x_i}$ from the previous generation is altered to yield a new vector $\vec{u_i}$ in the trial population by replacing individual components. The $n$th component $x_i^n$ is replaced with a constant probability $CR$ by the corresponding component $v_i^n$ of the mutated vector. $CR$ is the third DE constant called crossover. The $m$th component is always replaced, where $m$ is chosen randomly for each vector. This ensures that the trial population is always different from the previous generation. Therefore, a random integer number $m_i \in [0,d-1]$ is generated for each vector and a random real number $X_i^n \in ]0,1]$ is generated for each component $n$. The trial vector $\vec{u_i}$ is chosen according to

\begin{equation}
   u_i^n = \left\{
     \begin{array}{lcl}
       v_i^n & ; & X_i^n \leq CR \; \vee \; n=m_i \\
       x_i^n & ; & \textnormal{otherwise}
     \end{array}
   \right.
\nonumber
\end{equation}

(4) The experiment is evaluated for each individual vector $\vec{u_i}$ in the trial population. (5) The result is compared to the result of the corresponding vector $\vec{x_i}$ from the previous generation. The better performing vector is transferred to the new generation. Thus, each new generation outperforms both the previous generation and the trial population. (6) Repetition of steps (3)-(5) leads to a convergence towards the global optimum. 

However, experimental results are subject to technical noise. Furthermore, they may slightly drift due to variations of the environment (i.e. room temperature drifts). Therefore, the algorithm LILDE proposes two extensions to the original DE to account for noise and drift. 

\begin{itemize}
	\item The main modification introduces a limited lifetime for every individual vector $\vec{u_i}$. Each vector that survives a certain number of generations is considered to be outdated and is evaluated again. This avoids the accumulation of inferior individuals that accidentally yield a pseudo-superior result due to technical noise. Furthermore, the limited lifetime guarantees the continuous adaptation of the population to the drifting environment.
	\item The termination criteria proposed in the literature~\cite{Zielinski2008} are not appropriate in our case. Consequently, an alternative termination criterion is developed. We use the values of the objective function for the current generation. Our termination criterion is met when the set's standard deviation divided by its mean value falls below a termination threshold $T$. This provides invariance to the starting conditions and can be employed as long as near-zero objective function results are not expected.
\end{itemize}

Each new generation is tested for the termination criterion. (7) Finally the result of the optimization is given by the best performing vector of the last generation.

\begin{figure}[ht!]
		\includegraphics[width={6.4cm}]{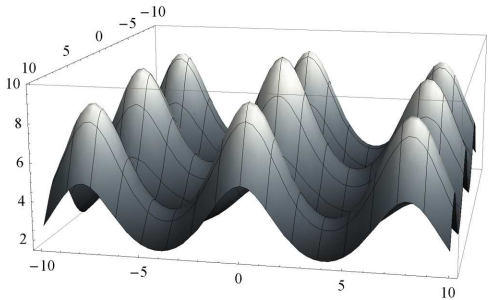}
		\caption{Analytical test function in two dimensions. We employ Ackley's function in multiple dimensions as an exemplary objective function.}
	\label{fig:ackley}
\end{figure}

In a first test the LILDE algorithm is applied to an exemplary analytic objective function to evaluate its performance. Here, we replace the experimental sequence by an evaluation of Ackley's function~\cite{Ackley1987} in a $d$-dimensional parameter space (see Fig.~\ref{fig:ackley}). We chose this function~\cite{AckleysFunction} because it features $3^{d}$ well-separated local maxima which are only $13\%$ lower than the central global maximum. 

\begin{figure}[ht!]
		\includegraphics[width={8.5cm}]{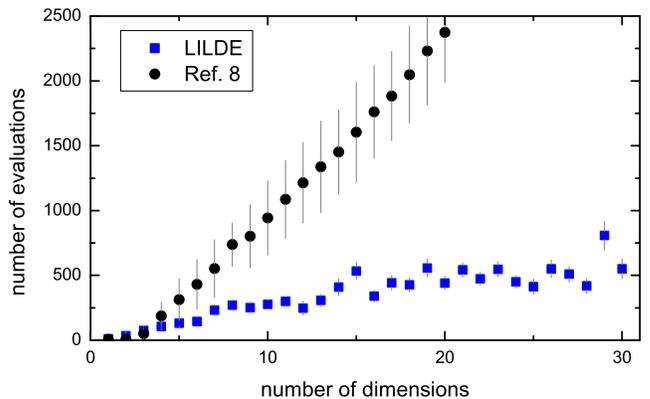}
		\caption{Scaling with the number of dimensions. The number of evaluations needed by the algorithms for convergence is depicted for increasing number of dimensions. The blue squares correspond to the LILDE algorithm and the black dots to the algorithm of Ref.~\cite{Rohringer2008}. Each data point is averaged over 50 runs of the simulation and the error bars indicate the standard deviation.}
	\label{fig:scalingTheo}
\end{figure}

Figure~\ref{fig:scalingTheo} shows the necessary number of evaluations for the optimization of Ackley's function for an increasing number of dimensions. We performed the optimization for various numbers of individuals per generation $N$ and chose the best performing case. The blue squares represent the mean number of evaluations needed by the LILDE algorithm - as reached with the optimal number of individuals. As a termination criterion, we demand all parameters of the best individual to reach the optimal value  within a $5\%$ range of the given parameter space, thus excluding all local maxima.  The error bars indicate the standard deviation within $50$ consecutive optimization runs. The black dots show the result obtained with the optimization algorithm from Ref.~\cite{Rohringer2008}. The optimal number of individuals per generation ranges for both algorithms between $6$ and $15$. Within this range, the optimization efficiency does not strongly depend on the exact choice. For the chosen objective function, we obtain an almost linear scaling with the number of dimensions for both algorithms. While the algorithm of Ref.~\cite{Rohringer2008} is efficient only for up to three dimensions, it is outperformed by LILDE by over a factor of five for higher dimensions.

\begin{figure}[h!tb]
		\includegraphics[width={8.5cm}]{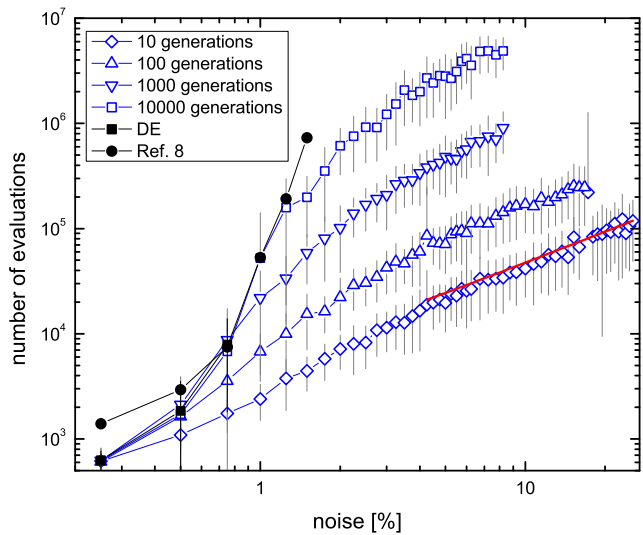}
		\caption{Performance for increasing noise on the objective function. For increasing noise, the DE algorithm (solid squares) and the algorithm of Ref.~\cite{Rohringer2008} (solid dots) quickly become inefficient. For reasonably short lifetimes, the performance is greatly improved (open symbols).}
	\label{fig:noise}
\end{figure}

A major challenge for optimization algorithms are drifts and noise on the objective function. We test the robustness to noise by adding shot-to-shot Gaussian noise to Ackley's function in ten dimensions.  Figure~\ref{fig:noise} shows the necessary number of evaluations depending on the amount of noise on the objective function. The standard deviation of the noise is increased from zero to $25\%$ of the corresponding function value. The figure shows the result of an optimization with 15 individuals per generation. The solid symbols represent the results of the algorithm of Ref.~\cite{Rohringer2008} (solid dots) and the DE algorithm with unlimited lifetime of the individuals (solid squares). The performance of both algorithms is drastically reduced for increasing noise. Both algorithms fail to reliably converge within $7.5\,$million evaluations for noise exceeding $1.5\%$ and $0.75\%$, respectively. However, with limited lifetime (open symbols) LILDE is robust to experimental noise. Shorter lifetime is in general superior, since the effort for remeasurement remains negligible. We recommend a lifetime of 10 generations, since the scaling is sufficiently improved and a remeasurement effort of $10\%$ is acceptable. For a sufficiently short lifetime, the number of evaluations scales linearly with the amount of noise (red line). Thereby, LILDE scales better than the literature approach~\cite{Jin2005,Das2005}, where noise is reduced by resampling and averaging (quadratic scaling). Thus, the presented analysis proves that LILDE is a well-suited algorithm for optimization tasks in a noisy environment.

In our experimental apparatus~\cite{Joellenbeck2011}, we apply the optimization algorithm to maximize the output of a source of trapped ultracold \Rb atoms. A precooled atom beam loads a three-dimensional magneto-optical trap (MOT). Subsequently, the atoms are cooled in an optical molasses, optically pumped to a magnetically trappable spin state and finally captured in a magnetic quadrupole field. The optimization objective is to capture as many atoms as possible in the magnetic quadrupole trap.

All magnetic fields for the MOT, the optical pumping, and the magnetic trap are generated by a mesoscopic atom chip. This atom chip consists of a structure of millimeter-scale wires. In the following, we employ a total of nine wires. The MOT and the optical molasses are operated with three back-reflected mutually orthogonal laser beams which are red-detuned to the \Rb cooling transition. A fourth laser beam is used for optical pumping. 

We optimize two sets of parameters which belong to two independent stages in the experimental sequence. The first stage, the loading of the MOT, is controlled with a set of nine parameters: eight parameters control the currents through the nine wires (two wires are connected in series) and one parameter adjusts the detuning of the MOT laser frequency. The second stage comprises 21 parameters. It consists of the following steps to transfer the captured atoms to the quadrupole trap.

\begin{itemize}
	\item The atoms are pulled closer to the atom chip by dynamically adjusting the MOT (8 currents, 1 laser frequency).
	\item The atoms are further cooled in a two-part optical molasses without magnetic fields. The MOT laser detuning is linearly ramped from a first to a second frequency. Afterwards, it is fixed to a third value (3 laser frequencies).
	\item Two wires provide a magnetic field for the optical pumping (2 currents).
	\item The atoms are loaded into a quadrupole trap (7 currents).
\end{itemize}
All parameters in the two sets are correlated such that the sets cannot be divided into independent subsets.

The proposed algorithm is implemented in LabVIEW. The software controls the whole apparatus and evaluates the achieved number of atoms. A real-time capable control system based on a Field-Programmable Gate Array outputs all relevant parameters via digital-to-analog converters. The number of captured atoms is measured by fluorescence imaging on a CCD camera. 

\begin{figure}[h!tb]
		\includegraphics[width={8.5cm}]{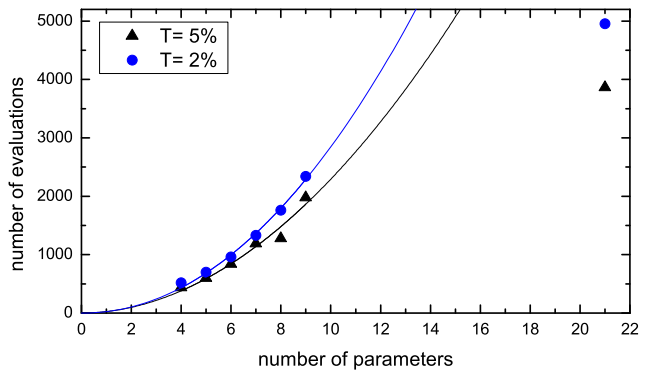}
		\caption{Scaling with the number of parameters. The number of experimental evaluations that are necessary to optimize a given number of parameters are shown. The black triangles and blue circles correspond to a termination threshold $T$ of $5\%$ and $2\,\%$, respectivley. The lines are quadratic fits to the data for up to 9 parameters.  The number of evaluations needed for set with 21 parameters falls below the quadratic trend, since a smaller number of vectors $N$ per generation is used. }
	\label{fig:scalingExp}
\end{figure}

Figure~\ref{fig:scalingExp} shows the number of measurements needed for the optimization of a given number of parameters. It comprises the two described sets of 9 and 21 parameters as well as smaller subsets of the 9 parameters. For each set, the evaluations needed with a termination threshold of both $5\,\%$ and $2\,\%$ are depicted. For the $2\,\%$ criterion, a cycle time of $3.5$~s results in total optimization durations of 2 hours 45 minutes (5 hours 45 minutes) for the case of 9 parameters (21 parameters).  The optimization was performed with the following DE constants: $F=0.9$, $CR=0.9$, and $E=0.5$. The number of vectors $N$ per generation is the $N=d\cdot 10$ for the measurements with $d \leq 9$. Here, we adopt recommended values from literature~\cite{Storn1995}. The solid lines are quadratic fits, presenting an almost quadratically increasing optimization effort. For a larger number of parameters, less vectors per dimension are needed. The 21-dimensional set was optimized using $N=d\cdot 4=84$, indicating that a scaling even better than quadratic can be reached. Thus, an optimal number of trapped atoms is reached which surpasses the result of the previous manual optimization in all presented cases.

In conclusion, we have shown that computer-based global optimization is a useful tool in experimental atom optics. We have presented an algorithm called LILDE which scales well with the number of optimized parameters and is specifically robust against experimental noise. Due to its broad applicability, the optimization with LILDE can be of great value in a wide spectrum of experimental tasks. 

Many experiments demand the simultaneous optimization with respect to conflicting objectives. In the future, the algorithm is easily extended to allow for such a multi-objective optimization~\cite{Babu2003}.

We acknowledge support from the Centre for Quantum Engineering and Space-Time Research QUEST and from the Deutsche Forschungsgemeinschaft (Research Training Group 1729). 

\bibliography{Lilde}

\begin{thebibliography}{19}
\bibitem [{Ack()}]{AckleysFunction}%
  \BibitemOpen
  \href@noop {} {\enquote {\bibinfo {title} {The function was parametrized
  according to \\ $\scriptstyle f(\vec{x_i})=\frac{1}{0.3}\{10 \exp \left(
  -\frac{1}{200} \sqrt{\frac{1}{d} \sum_{j=0}^{d-1} {{x_i^j}^2}}\right) +\exp
  \left( \frac{1}{d} \sum_{j=0}^{d-1}{\cos \left( \frac{\pi}{4} x_i^j
  \right)}\right) -9.7\} $ with d being the number of dimensions and $x_i$
  ranging from -10 to 10.}}\ }\BibitemShut {NoStop}%
\end{thebibliography}

\end{document}